\documentclass[aps,preprint,amsmath,amssymb,showpacs,superscriptaddress]{revtex4}
\usepackage{graphicx}
\usepackage{dcolumn}
\usepackage{bm}

\begin{document}


\title[] {Synchronization in interdependent networks}

\author{Jaegon Um}
\affiliation{School of Physics, Korea Institute for Advanced Study,
Seoul 130-722, Korea}
\author{Petter Minnhagen}
\affiliation{Department of Physics, Ume{\aa} University, 
901 87 Ume\aa, Sweden}
\author{Beom Jun Kim}
\email{beomjun@skku.edu}
\affiliation{BK21 Physics Research Division and Department of Physics,
Sungkyunkwan University, Suwon 440-746, Korea}
\affiliation{Asia Pacific Center for Theoretical Physics, Pohang 790-784, Korea}
\date{\today}
\begin{abstract}
We explore the synchronization behavior in interdependent systems, 
where the one-dimensional (1D) network 
(the intranetwork coupling strength $J_{\rm I}$)
is ferromagnetically intercoupled (the strength $J$)
to the Watts-Strogatz (WS) small-world network 
(the intranetwork coupling strength $J_{\rm II}$). 
In the absence of the internetwork coupling ($J = 0$),  
the former network is well known not to exhibit the synchronized phase 
at any finite coupling strength, whereas the latter 
displays the mean-field transition.
Through an analytic approach based on the mean-field approximation, 
it is found that for the weakly coupled 1D network ($J_{\rm I} \ll 1$)
the increase of $J$ suppresses synchrony, because the nonsynchronized
1D network becomes a heavier burden for the synchronization process of
the WS network. 
As the coupling in the 1D network becomes stronger, 
it is revealed by the renormalization group (RG) argument that 
the synchronization is enhanced as $J_{\rm I}$ is increased, implying that
the more enhanced partial synchronization in the 1D network 
makes the burden lighter.
Extensive numerical simulations confirm these expected behaviors, while
exhibiting a reentrant behavior in the intermediate range of $J_{\rm I}$. 
The nonmonotonic change of the critical value of $J_{\rm II}$ is also compared
with the result from the numerical RG calculation.
\end{abstract}

\pacs{05.45.Xt,05.70.Jk,89.75.Hc}

\maketitle

\begin{quotation}
Synchronization phenomenon is the one of the most fascinating 
collective emergent behaviors abundantly found in natural and
artificial systems. The onset of synchronization occurs when the
differences of individual oscillators are overcome by the
strong coupling among elements.
We investigate the coupled system of two networks, one is synchronizable 
at finite coupling strength while the other is not, aiming to 
answer the question of what happens as the inter- and intra-couplings 
are varied.
For the weak intracouplings of the nonsynchronizable
network, both our analytic and numerical results show that the stronger
internetwork coupling hinders the synchronization in the 
synchronizable network since the nonsynchronized oscillators in the other
network work as heavier burdens for the oscillators in 
the synchronizable network to carry to synchronize.
On the other hand, as the intracoupling strength in the nonsynchronizable
network becomes larger, partially synchronized groups of more oscillators
are formed, which in turn help the oscillators in the synchronizable network 
to become unified as synchronized clusters. In the intermediate regime
where the intra- and inter-network couplings are in the same order
of magnitude, numerical results show a reentrant behavior in the
synchronization phase diagram.
\end{quotation}


\section{Introduction}
Synchronization as a collectively emergent phenomenon in complex systems 
has attracted much interest thanks to the abundance of examples 
in nature~\cite{intro1,intro2,intro3,Kuramoto}. 
In existing studies, it has been revealed that the topology 
of interaction plays an important role in synchronizability~\cite{synch,
synchER}. In particular, it is now well known that coupled phase oscillators 
described by the celebrated Kuramoto model~\cite{Kuramoto} 
exhibit various universality classes depending on dimensionality and
the topological structure of networks~\cite{synchER, synchWS, ott,
synchSF, lattice1,lattice2}.
It has also been found that the lower critical dimension for the 
frequency synchronization in the regular $d$-dimensional Kuramoto model
is $d=2$, while the corresponding lower critical dimension for the
phase synchronization transition with the spontaneous $O(2)$ 
symmetry breaking is $d=4$~\cite{lattice1, lattice2}. Moreover, numerical 
studies~\cite{lattice2} have implied that systems of $d>4$ belong to the 
mean-field (MF) universality class as shown in globally coupled oscillators. 
On the other hand, for the Kuramoto model in complex networks, various 
results have been reported: For random and small-world 
networks~\cite{ER, WS}, it has been observed that 
the MF transition exists~\cite{synchER, synchWS}, and for scale-free (SF) 
network with a power-law degree distribution $\rho \sim k^{-\gamma}$, where 
$k$ stands for degree, $\gamma$-dependent critical exponents have been 
found via a MF approximation and numerical investigations~\cite{synchSF}.

\begin{figure}[t]
\includegraphics[width=7.5cm,clip]{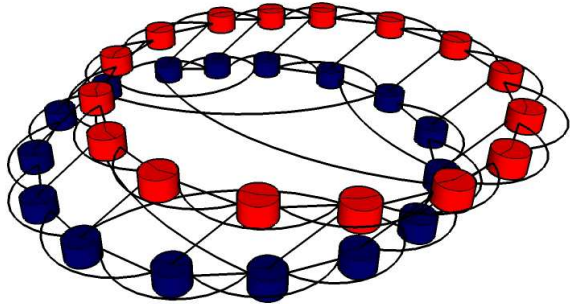}
\caption{\label{fig:net} (Color online) 
Two coupled interdependent networks: The one-dimensional regular network
(the upper one) 
with the periodic boundary condition is coupled to the Watts-Strogatz
small-world network (the lower one). Both networks have the average 
degrees four.}
\end{figure}

In the present work, we couple the one-dimensional (1D) network 
and the Watts-Strogatz (WS) small-world network~\cite{WS} 
(see Fig.~\ref{fig:net}), and 
investigate the synchronizability of Kuramoto oscillators in 
the composite two coupled networks.  Our research focus is put on 
the effect of internetwork coupling between the two networks 
belonging to different universalities, i.e., the MF universality for
the WS network and the absence of the synchronous phase for the 1D regular
network.  The physics of 
coupled interdependent networks are not only interesting in the
pure theoretical point of view, but it also can have practical applicability
since these interdependent network structures can be found ubiquitously. 
For example, electric power distribution in the power grid is 
strongly interwoven with the communication through Internet, and thus
spreading of failures in one network affects failures in other 
network~\cite{inter1, inter2}. 
This coupled system has been studied within the framework 
of percolation with the strength of internetwork coupling varied;
strong coupling between networks yields a first-order transition,
while in weakly coupled networks a giant cluster continuously vanishes 
at the critical point~\cite{inter3}. In Ref.~\onlinecite{jo}, the
epidemic spreading behavior has been studied in the coupled networks 
of the infection layer and the prevention layer.
We believe that the study
of the collective synchronization in interdependent networks
can also be an important realistic problem in a broader context: 
Imagine that each agent
in a social system has two different types of dynamic variables and that
the interaction of the one type of variable has different interaction
topology than the other variable.
We emphasize that our study of the synchronization in coupled networks 
is worthwhile because it could be applicable to investigate  
social collective behaviors in interdependent networks.   
Another interesting example of the interdependent system 
can be found in the neural network in the brain: The cortical region 
is coupled with the thalamus~\cite{inter4}, and the thalamocortical 
interactions might be interpreted as the intercoupling between
the cortical area and thalamus.

\section{Model}
We consider the composite system of two coupled networks 
(I and II) with equal number $N$ of oscillators for each.
The equations of motion for the Kuramoto model in the system are given by
\begin{widetext}
\begin{equation}
\dot{\phi}^{\rm I}_{j} =  \omega^{\rm I}_{j}-J\sin(\phi^{\rm I}_{j}-
\phi^{\rm II}_{j}) 
-J_{\rm I}\sum_{l}^{N}a^{\rm I}_{jl}\sin(\phi^{\rm I}_{j}-\phi^{\rm I}_{l}), 
\; \;
\dot{\phi}^{\rm II}_{j} =  \omega^{\rm II}_{j}-J\sin(\phi^{\rm II}_{j}-
\phi^{\rm I}_{j}) 
-J_{\rm II}\sum_{l}^{N}a^{\rm II}_{jl}\sin(\phi^{\rm II}_{j}-\phi^{\rm II}_{l}),
\label{full_eq}
\end{equation}
\end{widetext}
where $\phi^{\rm I (II)}_{j}$ is the phase of the $j$th oscillator 
and $\omega^{\rm I (II)}_{j}$ is its intrinsic frequency in the network 
I (II), assumed as an independent quenched random Gaussian variable 
with the unit variance. 
Throughout the paper, we denote the 1D regular network as 
the network I and the WS network as II, and use the average degree $\langle k
\rangle = 4$ for both. The adjacency matrix for the 1D network
has elements $a^{\rm I}_{jl}=\delta_{j,l\pm 1} + \delta_{j,l \pm,2}$, while  
$a^{\rm II}_{jl}$ is constructed following the small-world
network generation method: Each link is visited and
rewired at the probability $p$ (see Ref.~\cite{WS} for details).
In Eq.~(\ref{full_eq}), $J$ denotes the internetwork coupling strength between 
I and II, and $J_{\rm I(II)}$ stands for the intranetwork coupling for
the network I(II).  One may expect that when $J$ becomes large enough
the partial synchronization in the 1D regular network could be induced by 
the established ordering in the WS network although no spontaneous 
ordering exists in the pure 1D network. On the other hand, if $J_{\rm I}$
is not big enough, the internetwork coupling $J$ to the 
nonsynchronized 1D network could also make the WS network itself hard
to synchronize.  It is also possible that if  $J_{\rm I}$ is infinite, 
yielding fully synchronized 1D oscillators, 
synchronization is induced in the WS network even at $J_{\rm II}=0$. 
The interplay among these inter- and intra-network couplings can provide
rich phenomena, the understanding of which composes the main motivation of the
present study.

\section{Analytic Results}

\subsection{Mean-field analysis for $J=0$}
\label{subsec:mf}

When the two networks are decoupled, i.e., when $J=0$ in 
Eq.~(\ref{full_eq}), the network II should show the MF 
synchronization transition as reported in previous studies~\cite{synchWS}.
In this Section, we briefly review the MF theory 
of the Kuramoto model for the single WS network~\cite{ott, synchSF}. 
The equations of motion for the network II without $J$ is rewritten as 
\begin{equation}
\dot{\phi}^{\rm II}_{j} = \omega^{\rm II}_{j} -J_{\rm II} k^{\rm II}_{j} H_{j} 
\sin(\phi^{\rm II}_{j}- \theta_{j}), 
\label{short_eq}
\end{equation}    
where we have used $H_{j}e^{i\theta_{j}} \equiv  (1/k^{\rm II}_j)\sum_{l}^{N} 
a^{\rm II}_{jl} e^{i\phi^{\rm II}_l}$ with $k^{\rm II}_{j}$ 
being the degree of the $j$th node in II. In the spirit
of the MF approximation, we neglect the fluctuation and
substitute $H_{j}$ and $\theta_{j}$ by global variables $H$ and $\theta$, 
respectively, which yields the self-consistent equation for the 
order parameter $H$:
\begin{widetext}
\begin{equation}
H=\frac{1}{N \langle k_{\rm II} \rangle}\sum^{N}_{j} k^{\rm II}_{j} 
\sqrt{ 1- \left( \frac{\omega^{\rm II}_j}{k^{\rm II}_{j}HJ_{\rm II}} 
\right)^2 }
\Theta\left( 1- \frac{|\omega^{\rm II}_j|}{k^{\rm II}_{j}HJ_{\rm II}}
 \right),
\label{sc_eq1}
\end{equation}  
\end{widetext}
where $\Theta(x)$ is the Heaviside step function [$\Theta(x)=1 (0)$ for 
$x \geq 0$ ($ < 0$)]. In thermodynamic limit of $N \to \infty$, we
change the sum over oscillators to the sum over different degrees, which 
gives us  
\begin{equation}
H=\frac{1}{\langle k_{\rm II} \rangle}
\sum_{k_{\rm II}}\rho(k_{\rm II}) k_{\rm II} u(k_{\rm II}HJ_{\rm II}), 
\label{sc_eq2}
\end{equation}
where $\rho(k_{\rm II})$ is the degree distribution function and 
\begin{equation}
u(x) \equiv \int^{x}_{-x} d\omega_{\rm II} g(\omega_{\rm II})\sqrt{1-\omega_{\rm II}^2/x^2} 
\label{u_eq1}
\end{equation}
with $g(\omega_{\rm II}) = \exp(-\omega_{\rm II}^2/2\sigma_{\rm II}^2)/\sqrt{2\pi}\sigma_{\rm II}$. 
Note that near the critical point where $H$ becomes vanishingly small, 
Eq.~(\ref{u_eq1}) is expanded in the form 
$u \approx \frac{\pi}{2}g(0)k_{\rm II}HJ_{\rm II} +\frac{\pi}{16} g''(0) 
[k_{\rm II}HJ_{\rm II}]^3$, which leads to
\begin{equation}
H= \frac{\langle k^2_{\rm II} \rangle \pi }{2 \langle k_{\rm II} \rangle }
g(0)J_{\rm II}H +\frac{\langle k^4_{\rm II} \rangle \pi }{16 \langle k_{\rm II} 
\rangle } g''(0)J^{3}_{\rm II}H^{3}.
\label{sc_eq3}
\end{equation}
Here, it is to be noted that since the network II is the WS network with 
the exponential degree distribution, both 
$\langle k^{2}_{\rm II} \rangle$ and $\langle k^{4}_{\rm II} \rangle$ 
have well-defined finite values. It is then straightforward to 
get the critical point $J^{\rm II}_{c}= \frac{2\langle k_{\rm II} 
\rangle \sigma_{\rm II} \sqrt{2}}{\langle k^2_{\rm II} \rangle \sqrt{\pi}}$ and 
the critical exponent $\beta=1/2$. Moreover, introducing a sample-to-sample 
fluctuation~\cite{synchSF} to the right-hand side of Eq.~(\ref{sc_eq3}), 
we obtain the finite-size scaling (FSS) exponent $\bar{\nu}=5/2$.

\subsection{Mean-field analysis for $J_{\rm I} \ll 1$}
\label{subsec:weak}

We next turn our attention to the coupled system ($J \neq 0$),
and first consider the case of vanishingly small intranetwork
coupling for I. As $J_{\rm I} \to 0$, dynamics in the network I 
is simply governed by $\dot{\phi_{\rm I}} = \omega_{\rm I} -
J\sin(\phi_{\rm I}-\phi_{\rm II})$ with the site index omitted for convenience. 
For $\langle \omega_{\rm I} \rangle = \langle \omega_{\rm II} \rangle =0$, 
running oscillators in I with $\dot{\phi_{\rm I}} \neq 0$ may  
hinder their connected counterpart oscillators in II from entering into
the global entrainment. It has also been found that contributions of detrained 
oscillators to the synchronization order parameter are negligible within
the MF theory~\cite{ott, synchSF}. Accordingly, one can make the plausible 
assumption that entrained oscillators ($\dot{\phi_{\rm I}} = 0$) in I and their 
corresponding oscillators in II  mainly contribute to the synchronization. 
We then write the equations of motion for the entrained oscillators in II as 
\begin{equation}
\dot{\phi}_{\rm II} = \omega_{\rm II} +\tilde{\omega}_{\rm I} -J_{\rm II} 
k_{\rm II} H \sin(\phi_{\rm II}- \theta),
\label{short_eq2}
\end{equation} 
where $\tilde{\omega}_{\rm I} \equiv J \sin(\phi_{\rm I}-\phi_{\rm II})$ 
with $|\tilde{\omega}_{\rm I}| \leq J$, and  
the MF approximation ($H_j,\theta_j)  \rightarrow (H,\theta)$ 
has been made in the assumption that 
the internetwork coupling does not change the universality
of II since there exists no ordering in I for $J_{\rm I} \rightarrow 0$.
Consequently, the self-consistent equation reads 
$H=\left( N' \langle k_{\rm II} \rangle \right)^{-1}
\sum_{j}^{N'}k^{\rm II}_{j} \sqrt{ 1-f_{j}^{2}}
\Theta(1-|f_{j}|)$, where $f_{j} \equiv (\omega^{\rm II}_{j} + 
\tilde{\omega}^{\rm I}_{j})/k^{\rm II}_{j}HJ_{\rm II}$ and   
$N' \equiv  N\int_{-J}^{J} d\omega_{\rm I} g(\omega_{\rm I})$. 
In thermodynamic limit of $N \to \infty$, 
we again meet the form of $H=(1/\langle k_{\rm II} \rangle) 
\sum_{k_{\rm II}}\rho(k_{\rm II}) k_{\rm II}u(x)$ with 
\begin{widetext}
\begin{equation}
u(x) \equiv \int_{-J}^{J} d\omega_{\rm I}
\tilde{g}(\omega_{\rm I})
\int^{x-\omega_{\rm I}}_{-x-\omega_{\rm I}} 
d\omega_{\rm II} g(\omega_{\rm II})
\sqrt{1- \left( \frac{\omega_{\rm II} 
+\omega_{\rm I}}{x} \right)^2},
\label{u_eq2}
\end{equation}  
where $x=k_{\rm II}HJ_{\rm II}$ and
$\tilde{g}(\omega_{\rm I})= \mathcal{N}^{-1} 
e^{-\omega^{2}_{\rm I}/2\sigma^{2}_{\rm I}}$ with the normalization constant
$ \mathcal{N}=\int_{-J}^{J} d\omega_{\rm I} 
e^{-\omega^{2}_{\rm I}/2\sigma^{2}_{\rm I}}$. Following the similar 
steps to those made for $J=0$,  we conclude that the synchronization transition
occurs at the critical value 
$J^{\rm II}_{c} = \frac{2\langle k_{\rm II} \rangle}
{A \langle k^{2}_{\rm II} \rangle \pi}$  
with a constant 
\begin{equation}
A \equiv \left(\sqrt{2\pi \sigma^{2}_{\rm II}} \mathcal{N} \right)^{-1}
\int_{-J}^{J} d\omega_{\rm I}  
\exp\left[-\omega^{2}_{\rm I} 
\left( \frac{2\sigma^{2}_{\rm I}\sigma^{2}_{\rm II} } 
{\sigma^{2}_{\rm I} + \sigma^{2}_{\rm II} } \right)^{-1} \right].
\label{A_eq}
\end{equation}
\end{widetext}
We note from the expression that as $J$ is increased, 
$J_{c}^{\rm II}$ also increases since $A$ is decreased due to 
the fact that $\frac{\sigma^{2}_{\rm I}\sigma^{2}_{\rm II} }
{\sigma^{2}_{\rm I} + \sigma^{2}_{\rm II} } < \sigma^{2}_{\rm I}$. Since 
$\sigma_{\rm I}=\sigma_{\rm II}$,
one obtains $J_{c}^{\rm II}(J) = \frac{4 \langle k_{\rm II} \rangle 
\sigma_{\rm II}}{\langle k^{2}_{\rm II} \rangle \sqrt{\pi}}=
\sqrt{2}J_{c}^{\rm II}(0)$ with $J \to \infty$. 
Furthermore, it is expected that as $J_{\rm I}$ is increased (but still
$J_{\rm I} \ll 1$) $J^{\rm II}_c$ should decrease from the following
reasoning: 
When $J_{\rm I} \ll 1$, effective equations for oscillators in I can be 
written as $\dot{\phi_{\rm I}} = \omega'_{\rm I} -
J\sin(\phi_{\rm I}-\phi_{\rm II})$. Here, $\omega'_{\rm I}$ is a modified
frequency whose variance $\sigma'_{\rm I}$ becomes smaller than the bare 
value due to the attractive force activated by the existence of 
the intranetwork coupling.
We then conclude that $\sigma'_{\rm I} \equiv \sigma'_{\rm I}(J_{\rm I})$ is
a decreasing function of $J_{\rm I}$. 
Substituting $\sigma_{\rm I}$ by $\sigma'_{\rm I}$ in Eq.~(\ref{A_eq}), 
one notes that $ r(J_{\rm I}) \equiv\sigma^{2}_{\rm II}/ 
(\sigma'^{2}_{\rm I} + \sigma^{2}_{\rm II} )$ increases with $J_{\rm I}$,
and thus $A= \left( \sqrt{2\pi\sigma_{\rm II}^{2} }
\mathcal{N} \right)^{-1} \int_{-J}^{J} 
d\omega_{\rm I} \exp\left[ -\omega_{\rm I}^2/ 
\left( 2 r\sigma'^{2}_{\rm I}\right) \right] $ with  
$\mathcal{N} =\int_{-J}^{J} d\omega_{\rm I} 
\exp\left[ -\omega_{\rm I}^2/ \left( 2 \sigma'^{2}_{\rm I}\right) 
\right]$ becomes an increasing function (i.e.,  $J^{\rm II}_c$ becomes
a decreasing function) with respect to $J_{\rm I}$. 
In summary of this subsection,  when the 1D network is within the weakly 
coupled regime  (i.e., when $J_{\rm I}$ is sufficiently small), 
the synchronization of the WS network is enhanced  (i.e., $J^{\rm II}_c$ 
in reduced), as $J_{\rm I}/J$ is increased. In words, the stronger internetwork
coupling puts more  burden for the WS network to achieve its synchrony, while
the better synchrony in the 1D network helps the WS network to be better
synchronized via the internetwork coupling.

\subsection{Renormalization group approach for $J_{\rm I} > J$ }
\label{subsec:strong}
For a strong coupling regime of the 1D regular network,  
our MF equations in Sec.~\ref{subsec:weak} for $J_{\rm I} \to 0$ need
to be modified by employing the real-space renormalization-group (RG) 
formulation that has been developed in the 1D systems~\cite{RG}. 
In this RG approach applied for 1D regular networks, 
strong bonds form clusters of entrained oscillators, 
while fast moving oscillators are decimated, interrupting the development of 
a giant synchronized cluster. In our system, 
it is expected that the strong coupling in the 1D regular network
should induce synchronized clusters not only in the 1D regular network, 
but also in the WS network coupled to it via the internetwork coupling.

For $J_{\rm I} >J$, oscillators in the network I are governed 
mainly by intranetwork coupling rather than by internetwork 
coupling, which allows us to apply the RG approach to the 1D network
with respect to $\omega_{\rm I}$ and $J_{\rm I}$: For 
$|\omega_{\rm I}| > J_{\rm I} >J$, 
$\omega_{\rm I}$ becomes relevant, whereas for $J_{\rm I}>|\omega_{\rm I}|>J$
and $J_{\rm I} > J > |\omega_{\rm I}|$, $J_{\rm I}$ becomes relevant.
Applying the RG approach, fast moving oscillators 
having $|\omega_{\rm I}|>J_{\rm I}$ are removed from the 1D regular 
network together with their bonds, yielding fragmentation in I (our numerical
RG calculations are summarized in Sec.~\ref{sec:numeric}). 
On the other hand, since the remaining bonds are strong enough,
oscillators left in a fragment form a synchronized cluster with the phase 
$\Phi^{\rm I}_{j}$ and the renormalized frequency $\Omega^{\rm I}_{j} 
=(1/m^{\rm I}_{j}) \sum_{i \in S^{\rm I}_{j}} \omega^{\rm I}_{i}$, 
where $m^{\rm I}_{j} \equiv |S^{\rm I}_j|$ is a number of 
oscillators in the $j$th cluster $S^{\rm I}_{j}$.

The entrainment of oscillators into clusters 
in the network I tends to drive
oscillators in network II to cluster themselves in the same 
corresponding clusters because of the ferromagnetic intercouplings. 
For simplicity, we assume that the same number $N_{c}$
of clusters are generated both in I and II.
Within the framework of the MF approximation, equations
of motion are written as
\begin{widetext}
\begin{equation}
\dot{\Phi}^{\rm I}_{j} = \Omega^{\rm I}_{j} -J\sin(\Phi^{\rm I}_{j}
-\Phi^{\rm II}_{j}), \; \;
\dot{\Phi}^{\rm II}_{j}= \Omega^{\rm II}_{j} -J\sin(\Phi^{\rm II}_{j}
-\Phi^{\rm I}_{j}) - \langle k_{\rm II} \rangle J_{\rm II}H
\sin(\Phi^{\rm II}_{j} -\theta), 
\label{rg_eq}
\end{equation} 
\end{widetext}
where $\Omega^{\rm II}_{j} =(1/m^{\rm II}_{j}) \sum_{i \in S^{\rm II}_{j}} 
\omega^{\rm II}_{i}$ with $m^{\rm II}_{j}$ being the number of
oscillators in the $j$th cluster $S^{\rm II}_{j}$ in the network II. 
A set of oscillators $\{\phi^{\rm II}_{i}\}$ connected to $\Phi_{j}^{\rm I}$ 
are merged into $\Phi^{\rm II}_{j}$ except for ones with 
$|\omega_{\rm II}| >J$ since we only consider entrained clusters induced by the 
internetwork  coupling. $\langle k_{\rm II} \rangle$ in Eq.~(\ref{rg_eq})
also arises from $(1/m^{\rm II}_{j})\sum_{i \in S^{\rm II}_{j}} 
k^{\rm II}_{i} \approx \langle k_{\rm II} \rangle$. 
Now, it is straightforward to obtain 
$J_{c}^{\rm II}$ from Eq.~(\ref{rg_eq}) via the same procedure as done
for the case of $J_{\rm I}=0$, to yield
\begin{equation}
J^{\rm II}_{c} = \frac{2\sqrt{2\bar{\sigma}^{2}_{\rm I} 
+2\bar{\sigma}^{2}_{\rm II} } }{\langle k_{\rm II} \rangle \sqrt{\pi}},
\label{jc_eq}
\end{equation}
where the approximation $\int_{-J}^{J} d\Omega_{\rm I} \approx \int_{-\infty}^{\infty} 
d\Omega_{\rm I}$ has been made from the assumption that $\bar{\sigma}_{\rm I} 
< J$. Here, $\bar{\sigma}^{2}_{\rm I (II)}$ is the variance for 
$\Omega_{\rm I (II)}$ within the Gaussian approximation, that is, 
$g(\Omega_{\rm I(II)})= e^{-\Omega^{2}_{\rm I(II)} 
/2\bar{\sigma}_{\rm I(II)}^{2} }/\sqrt{2\pi\bar{\sigma}^{2}_{\rm I(II)}}$. 
Note that $\bar{\sigma}_{\rm I} \sim \sigma_{\rm I}^{*}/
\sqrt{\bar{m}_{\rm I}}$ and $\bar{\sigma}_{\rm II} 
\sim \sigma^{*}_{\rm II}/\sqrt{\bar{m}_{\rm II}}$, 
where $\bar{m}_{\rm I(II)}$ is the
average number of oscillators forming a cluster in I(II), given
by $\bar{m}_{\rm I(II)} = \left[ N \int_{-J_{\rm I}(-J)}^{J_{\rm I}(J)} 
d\omega_{\rm I(II)} g(\omega_{\rm I(II)}) \right]/N_{c}$, and 
$\sigma^{*}_{\rm I(II)}$ is obtained from 
$\sigma_{\rm I(II)}^{*2} =\int_{-J_{\rm I}(-J)}^{J_{\rm I}(J)} 
d\omega_{\rm I(II)} \omega^{2}_{\rm I(II)} \tilde{g}(\omega_{\rm I(II)})$ 
since the frequency of 
oscillators which consist of clusters in I(II) 
must be less than $J_{\rm I}$ ($J$).
For sufficiently large $J$ and $J_{\rm I}$ yielding 
$\sigma^{*}_{\rm I(II)} \sim 1$ and $\bar{m}_{\rm I(II)} \sim N/N_{c}$,
it is obtained that $J_{c}^{\rm II} \sim \sqrt{N_{c}/N }$, 
implying that the synchronization onset decreases with $J_{\rm I}$ 
since the number of clusters should be the decreasing function 
with large $J_{\rm I}$;
only at $J_{\rm I} \to \infty$, $N_{c}/N$ becomes close to $1/N$, yielding 
$J_{c}^{\rm II} =0$ in thermodynamic limit. 

We have above investigated the synchronization onset as a function of
$J$ and $J_{\rm I}$ via the MF approximation. It has been found that
for weakly coupled 1D oscillators, the internetwork coupling increases the
synchronization onset, while increasing $J_{\rm I}$ enhances the 
synchronizability. The role of the strong intranetwork coupling in the
1D regular network has been revealed through the RG approach; clustering 
in the 1D network makes 
the synchronization easier. 

For FSS exponent for $J_{\rm I}>0$, 
$\bar{\nu}=5/2$ still holds since the number of samples, i.e., $N'$ or $N_{c}$ 
is proportional to the system size $N$ linearly:  
The density $N_{c}/N$ of clusters only depends on $J_{\rm I}$ since 
$\omega_{\rm I}$ is a random variable, yielding $N_{c} \sim N$. 
 
\section{Numerical results}
\label{sec:numeric}

\begin{figure}[t]
\includegraphics[width=6.5cm,clip]{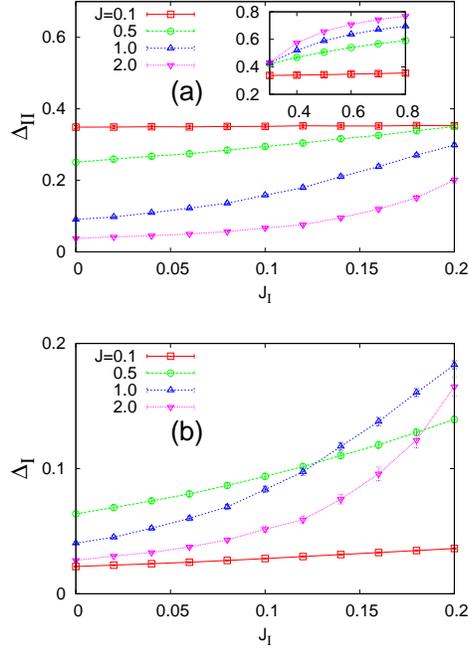}
\caption{\label{fig:delta} (Color online) 
Order parameters obtained through the numerical integrations 
of Eq.~(\ref{full_eq}):
(a) for the WS network ($\Delta_{\rm II}$) and
(b) for the 1D regular network ($\Delta_{\rm I}$) versus
the intranetwork coupling strength in the 1D network $J_{\rm I}$, 
at various values of internetwork coupling $J$ 
(the intranetwork coupling $J_{\rm II}$ in the WS network is set to 0.7
and the network size $N=12800$ for both I and II). For sufficiently small
values of $J_{\rm I}$, the synchrony in II is suppressed (enhanced)
as $J (J_{\rm I})$ is increased, as shown Sec.~\ref{subsec:weak} via 
the MF approach. 
Inset in (a): Further increase of $J_{\rm I}$ enhances 
the synchronization in II. The internetwork coupling $J$ plays a positive
role for the synchronization in II, differently from the weak-coupling
regime: As $J$ is increased at a fixed $J_{\rm I}$, $\Delta_{\rm II}$
is increased.
(b) shows that for small values
of $J_{\rm I}$ the increase of
internetwork coupling first enhances synchrony in I and then reduces it
as $J$ is increased further. }
\end{figure}

In order to validate our MF predictions, we perform numerical integrations
of Eq.~(\ref{full_eq}), with the network II constructed
via the WS model~\cite{WS} at the rewiring rate $p=0.5$. 
After achieving the steady state, we measure the time-averaged
order parameter
\begin{equation}
\Delta_{\rm I(II)} \equiv \frac{1}{T} \int_{T_{0}}^{T_{0}+T} dt \left < 
\frac{1}{N}\left| \sum_{j}^{N} e^{-i \phi_{j}^{\rm I(II)}
\left(t \right) }\right| \right >,
\end{equation} 
where $\langle \dots \rangle$ stands for the average over
different realizations of frequencies and networks.

First, we compute order parameters when the 1D network is in the
weakly coupled regime (small $J_{\rm I}$) at fixed $J_{\rm II}$.
In Fig.~\ref{fig:delta}, we exhibit $\Delta_{\rm II}$ and $\Delta_{\rm I}$ 
as functions of $J_{\rm I}$ in (a) and (b), respectively. 
It is observed that as the internetwork coupling $J$ is increased, 
the order parameters $\Delta_{\rm I}$ and $\Delta_{\rm II}$ 
show a decreasing tendency, except for
very small values of $J$ for $\Delta_{\rm I}$. The nonmonotonic change
of $\Delta_{\rm I}$ with respect to $J$ is not 
surprising, since $\Delta_{\rm I} \approx 0$ as $J \rightarrow 0$
[see the curve for $J = 0.1$ in Fig.~\ref{fig:delta}(b)].

\begin{figure}[t]
\includegraphics[width=6.5cm,clip]{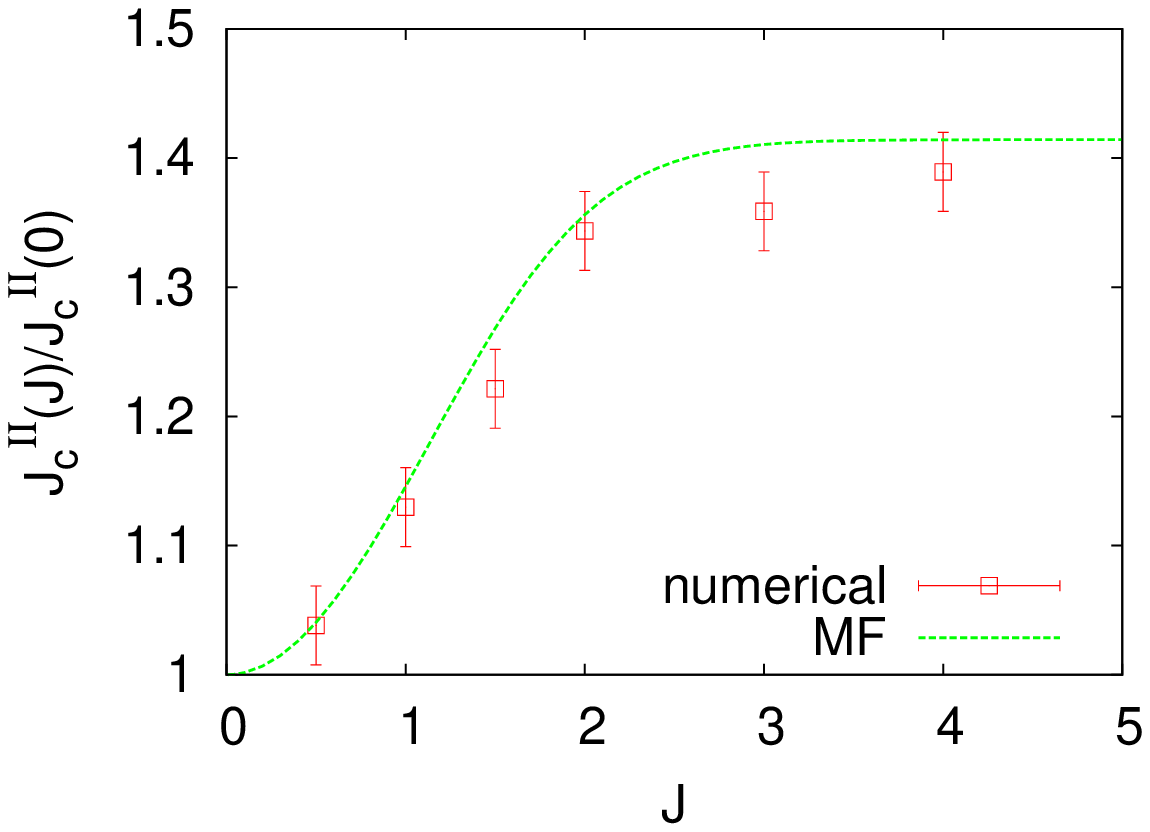}
\caption{\label{fig:mf} (Color online) 
Ratio $J_{c}^{\rm II}(J)/J_{c}^{\rm II}(0)$ as a function of $J$ 
at $J_{\rm I}=0$. Points are from  the numerical integrations of 
Eq.~(\ref{full_eq}) combined with the finite-size scalings in Eq.~(\ref{eq:FSS}), 
while the line denotes the MF prediction made in Sec.~\ref{subsec:weak}.    
}
\end{figure}

For $J_{\rm I}=0$, the synchronization onset $J_{c}^{\rm II}$ is evaluated
using the FSS of the form 
\begin{equation}
\label{eq:FSS}
\Delta_{\rm II} = N^{-\beta/\bar{\nu}}
f\bigl( ( J_{\rm II}-J_{c}^{\rm II}) N^{1/\bar{\nu}} \bigr)
\end{equation}
with $\beta=1/2$ 
and $\bar{\nu}=5/2$, resulting in $J_{c}^{\rm II}(J) \approx$
0.66, 0.68, 0.74, 0.8, 0.88, 0.89 and 0.91 
at $J=$ 0, 0.5, 1.0, 1.5, 2.0, 3.0 and 4.0, 
respectively. In Fig.~\ref{fig:mf}, it is observed that the ratio 
$J_{c}^{\rm II}(J)/J_{c}^{\rm II}(0)$ obtained numerically
follows the MF prediction in Sec.~\ref{subsec:weak}, given by
$J_{c}^{\rm II}(J)/J_{c}^{\rm II}(0) = \int_{0}^{J} dx e^{-x^2/2} 
/ \int_{0}^{J} dx e^{-x^2} $ with $\sigma_{\rm I}=\sigma_{\rm II}=1$. 
It is also found that $J_c^{\rm II}$ saturates 
at large $J$:  $J_{c}^{\rm II}(4.0)/J_{c}^{\rm II}(0) \approx 1.4$,
which is consistent with the MF value, 
$J_{c}^{\rm II}(J \to \infty)/J_{c}^{\rm II}(J = 0) 
 = \sqrt{2}$. 
Also in the weakly-coupled regime of small $J_{\rm I} (> 0)$, it is again
observed that $J_{c}^{\rm II}$ is increased as $J$ is increased, implying that
the stronger internetwork coupling  worsens the synchronizability of II, in
agreement with the finding in Sec.~\ref{subsec:weak}.  Another MF prediction in
Sec.~\ref{subsec:weak} that the increase of $J_{\rm I}$ enhances the synchrony
in II is clearly confirmed in Fig.~\ref{fig:delta}(a): At fixed $J$, the order
parameter for II is an increasing function of $J_{\rm I}$.  
For larger $J_{\rm I}$, the $J$-dependence of $\Delta_{\rm II}$ 
shown in the inset for Fig.~\ref{fig:delta}(a) can be interpreted as follows:  
More oscillators should be
involved in synchronization with stronger internetwork coupling in I, 
and thus the order parameter increases as $J$ is increased, which may be implying that
$J_{c}^{\rm II}$ is decreasing with $J$ at large $J_{\rm I}$ in contrast to the
case for the weak coupling regime of small $J_{\rm I}$. Consequently,
we conclude that the role of the internetwork coupling $J$ is reversed
in the weak and the strong coupling regimes: For the weakly (strongly) 
coupled 1D network, internetwork coupling worsens (enhances)
the synchrony in the WS network.

\begin{figure}
\includegraphics[width=7cm,clip]{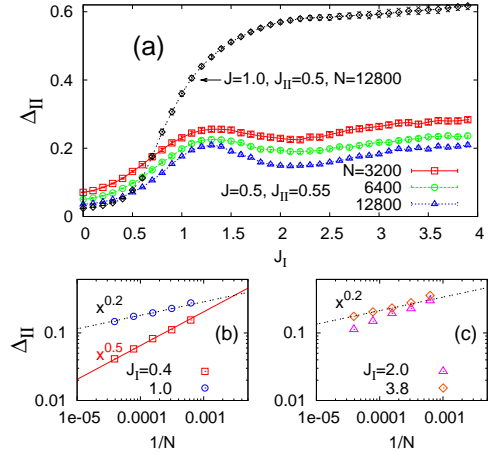}
\caption{\label{fig:reent} (Color online)
(a) $\Delta_{\rm II}$ versus $J_{\rm I}$ at $J=0.5$ and 
$J_{\rm II}=0.55$ for various system sizes. 
Nonmonotonic change of $\Delta_{\rm II}$ is observed at all system sizes.
$N$-dependence of $\Delta_{\rm II}$
for (b) $J_{\rm I}=$0.4, 1.0 and (c) 2.0, and 3.8 at $J=0.5$ and
$J_{\rm II}=0.55$.  The upper (lower) line is for $N^{-0.2}$ ($N^{-0.5})$. 
The MF-like scaling $N^{-0.2}$ is found at $J_{\rm I} \approx 1$ and at
 $J_{\rm I} \approx 4$, indicating that the system undergoes
MF transition twice as $J_{\rm I}$ is varied.
In (a), for comparison, $\Delta_{\rm II}$ for $J=1.0$ and $J_{\rm II}=0.5$
(the curve at the top) is shown to be monotonic.}
\end{figure}  

Very interestingly, a reentrant behavior of 
the order parameter as a function of $J_{\rm I}$ is
observed at certain values of $J$.  For example, at $J=0.5$ and 
$J_{\rm II}=0.55$, nonmonotonic behaviors of $\Delta_{\rm II}$ 
are displayed in Fig.~\ref{fig:reent} (a):
$\Delta_{\rm II}(J_{\rm I})$ increases with $J_{\rm I}$
in accord with the prediction in Sec.~\ref{subsec:weak}, 
begins to decreases at around $J_{\rm I} \approx 1.2$, and 
finally increases again. The increase of $\Delta_{\rm II}$ at 
large values of $J_{\rm I}$ is consistent with the calculation made
in Sec.~\ref{subsec:strong}. In Fig.~\ref{fig:reent} (a), also shown 
is that the nonmonotonic behavior does not change much with 
the system size. We observe that this reentrance is seen only in the
limited range of the internetwork coupling, and disappears at lager 
$J$, as displayed in Fig.~\ref{fig:reent} (a) (see the upper most curve for $J=1.0$). 
In Fig.~\ref{fig:reent} (b) and (c), we exhibit scaling behaviors of 
$\Delta_{\rm II}$ with $N$. What is found is that MF-like critical behavior,
$\Delta_{\rm II} \sim N^{-\beta/\bar{\nu}}$ with $\beta/\bar{\nu}=1/5$,  
is seen both at $J_{\rm I} \approx 1$ and $J_{\rm I} \approx 4$, 
whereas $\Delta_{\rm II}$ decays more rapidly with $N$
at other values [e.g., $J_{\rm I}=0.4$ and 2.0 in Fig.~\ref{fig:reent}(b) and (c)].
We believe that this observation clearly indicates that 
as $J_{\rm I}$ is increased the curve of $\Delta_{\rm II}
(J_{\rm I})$ touches phase boundary twice, first at $J_{\rm I} \approx 1$, 
and later at $J_{\rm I} \approx 4$, implying the existence of a
reentrant transition. 

\begin{figure}
\includegraphics[width=7cm,clip]{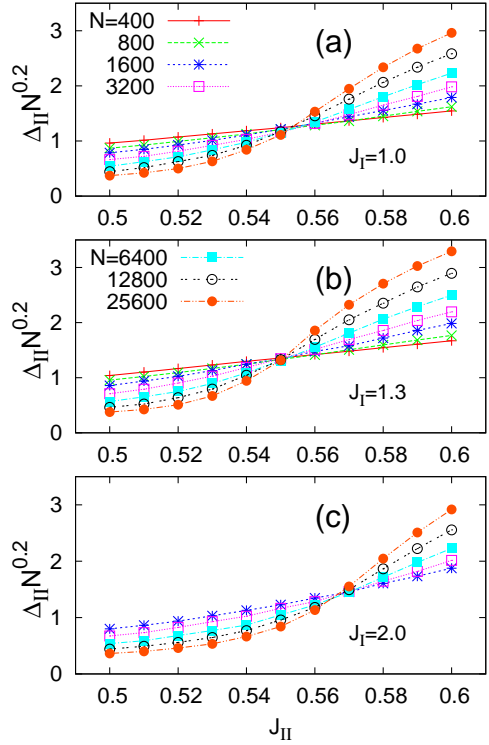}
\caption{\label{fig:fss} (Color online) 
Finite-size scaling [see Eq.~(\ref{eq:FSS})] 
for $\Delta_{\rm II}$ at $J=0.5$ and $J_{\rm I}=$ (a) 1.0,
(b) 1.3, and (c) 2.0. Crossing point first decreases and then increases,
indicating that there exists a reentrant transition. } 
\end{figure}  

To concrete our conclusion, we perform the FSS for $\Delta_{\rm II}$ as
a function of $J_{\rm II}$ with varying $J_{\rm I}$, as shown in
Fig.~\ref{fig:fss}. Since $N^{\beta/\bar{\nu}} \Delta_{\rm II} = 
f(0)$ at $J_{\rm II}=J_{c}^{\rm II}$ from Eq.~(\ref{eq:FSS}), 
crossings shown in Fig.~\ref{fig:fss} clearly manifest the reentrance behavior
of $J_{c}^{\rm II}$: The crossing point decreases as $J_{\rm I}$ is increased 
from 1.0 to 1.3, and then increases as $J_{\rm I}$ is increased to 2.0.
We report $J_{c}^{\rm II}$ obtained from FSS for $J=0.5$ in Fig.~\ref{fig:phd}:
$J_{c}^{\rm II}(J_{\rm I})$ first decreases from $J_{\rm I}=0$ as
predicted in Sec.{\ref{subsec:weak}, and makes an up turn before eventually
decreases again for larger $J_{\rm I}$, consistent with  Fig.~\ref{fig:reent}.     
Here, we emphasize that the reentrance behavior of the onset 
develops at $J_{\rm I}>J$.

\begin{figure}
\includegraphics[width=7cm,clip]{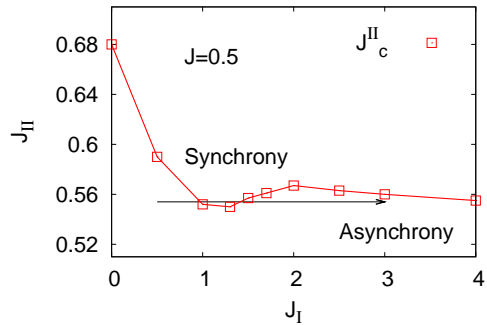}
\caption{\label{fig:phd} (Color online) Phase boundary at $J=0.5$ separating the synchronous (the
upper region) and the asynchronous (the lower region) phases. Points are
obtained from the FSS of $\Delta_{\rm II}$ [see Fig.~\ref{fig:fss}].
As $J_{\rm I}$ is increased from below at $J_{\rm II} \approx 0.56$, 
the system starts from asynchronous phase, enters the 
synchronous phase, and then reenters back to the asynchronous phase as noted
by the horizontal arrow. 
}
\end{figure} 

The hump structure of the synchronization onset at $J_{\rm I}>J$
in Fig.~\ref{fig:phd} can be understood qualitatively 
by recalling the MF theory for $J_{c}^{\rm II}$ at
$J_{\rm I}>J$ [see Eq.~(\ref{jc_eq})], given by a function of fluctuations of
frequency in clusters, 
$J_{c}^{\rm II}(J_{\rm I}) \sim \sqrt{\bar{\sigma}^{2}_{\rm I} +
\bar{\sigma}^{2}_{\rm II} }$. Rewriting Eq.~(\ref{jc_eq}) as
\begin{equation}
J_{c}^{\rm II}(J_{\rm I}) \sim \sqrt{\left(N_{c}/N \right)
\left(\mathcal{N}_{\rm I}^{-1}\sigma^{*2}_{\rm I} +
\mathcal{N}_{\rm II}^{-1}\sigma^{*2}_{\rm II} \right)}, 
\label{jc_eq2}
\end{equation}
with $\sigma^{*2}_{\rm I(II)}=\mathcal{N}^{-1}_{\rm I(II)}
\int_{-J_{\rm I}(-J)}^{J_{\rm I}(J)} 
d\omega_{\rm I(II)} \omega^{2}_{\rm I(II)}g(\omega_{\rm I(II)})$ and
$\mathcal{N}_{\rm I(II)} = \int_{-J_{\rm I}(-J)}^{J_{\rm I}(J)} 
d\omega_{\rm I(II)}g(\omega_{\rm I(II)})$,
one can find ingredients that determine the onset:
For fixed $J$, $\mathcal{N}_{\rm II}^{-1}\sigma^{*2}_{\rm II}$ is constant,
while $\mathcal{N}_{\rm I}^{-1}\sigma^{*2}_{\rm I}$ is an increasing
function of $J_{\rm I}$. It is expected that as $J_{\rm I}$ is
increased, $N_{c}$ increases at small $J_{\rm I}$ since clusters are created 
rather than merged. For large $J_{\rm I}$, on the other hand, 
$N_c$ should decrease as distinct clusters are merged.    

We finally perform numerical RG analysis for the 1D regular network 
as described in Sec.~\ref{subsec:strong}:
Initially, the Gaussian frequency is distributed for a 1D network of
the size $N$ up to $10^6$,
and oscillators with $|\omega_{\rm I}| >J_{\rm I}$ are 
removed together with their bonds. Collecting fragments of the network, 
we calculate $N_{c}$ as a function of $J_{\rm I}$. From the 
numerical RG analysis, we find that $N_{c}/N$ as a function of
$J_{\rm I}$ exhibits a well-defined peak near $J_{\rm I} \approx 0.45$
as seen in Fig.~\ref{fig:rg}.
This allows us to expect that $J_c^{\rm II}(J_{\rm I})$ 
has a peak at $J^{*}_{\rm I}$ slightly larger than $0.45$ 
since $\sigma^*_{\rm I}$ in the right-hand side of Eq.~(\ref{jc_eq2}) 
is also an increasing function of $J_{\rm I}$.
Since the expression for $J_{c}^{\rm II}$ in 
Eq.~(\ref{jc_eq2}) is valid only for $J_{\rm I}>J$, 
for larger $J$ than $J^{*}_{\rm I}$, 
the reentrance behavior disappears as $J$ is increased further,
as seen in Fig.~\ref{fig:reent}.
Although the values of $J_{\rm I}$ where the hump structure develops
are different from those from the numerical investigation, 
the MF result made in Eq.~(\ref{jc_eq2}) gives us a qualitatively correct
prediction. It indicates that when $N_{c}$ increases with increasing
$J_{\rm I}$, the synchronization in the WS network becomes worse since
1D clusters with different frequencies would act as burdens,
while the merging of clusters 
yields the better synchronization for sufficiently large $J_{\rm I}$.     

\begin{figure}
\includegraphics[width=7cm,clip]{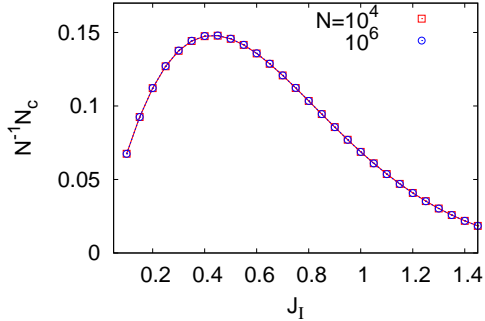}
\caption{\label{fig:rg} (Color online) Numerical RG calculation of $N_{c}/N$ as
a function of $J_{\rm I}$ in the 1D regular network with $\langle k \rangle=4$.
One thousand different sets of intrinsic frequencies are used for the average. }
\end{figure} 

\section{Summary}
We have investigated synchronization phenomenon in the interdependent
two coupled networks; one is the 1D regular network and the other
is the WS small-world network. Both the mean-field approximation and the
numerical simulations have shown that the effect of the internetwork coupling
is two folds: it suppresses the synchronization in the WS network 
when the 1D network is in the weak-coupling regime, while it enhances
the synchronization in the strong-coupling regime. 
In comparison, the intranetwork coupling in the 1D network has been shown 
to always play a positive role in the synchronizability of the WS network.
In the intermediate range of the intranetwork coupling, the reentrant behavior
has been found numerically and explained within the MF scheme combined with
the numerical RG calculations.

\section*{Acknowledgment}
This work was supported by NAP of Korea Research
Council of Fundamental Science \& Technology.
B.J.K. and J.U. thank 
the members of Icelab at Ume{\aa} University for hospitality during
their visit, where this work was initiated.


\begin{thebibliography}{6}
\bibitem{intro1} A. T. Winfree, {\it The Geometry of Biological Time}
(Springer-Verlag, Berlin, 1980).
\bibitem{intro2} A. S. Pikovsky, M. Rosenblum, and J. Kurths, {\it 
Synchronization: A Universal Concept in Nonlinear Science} (Cambridge 
University Press, Cambridge, 2001).
\bibitem{intro3} S. Boccaletti, V. Latora, Y. Moreno, M. Chavez, and 
D.-U. Hwang, Phys. Rep. {\bf 424}, 175 (2006). 
\bibitem{Kuramoto} Y. Kuramoto, in {\it Proceedings of the International
Symposium on Mathematical Problems in Theoretical Physics}, edited by 
H. Araki (Springer, Berlin, 1975);
{\it Chemical Oscillations, Waves, and Turbulence} (Springer, Berlin, 1984).
\bibitem{synch} L. M. Pecora and T. L. Carroll, Phys. Rev. Lett. {\bf 80}, 
2109 (1998); M. Barahona and L. M. Pecora, {\it ibid.} {\bf 89}, 054101 (2002);
H. Hong, B. J. Kim, M. Y. Choi, and H. Park, Phys. Rev. E {\bf 69}, 067105 
(2004); S. M. Park and B. J. Kim, {\it ibid.} {\bf 74}, 026114 (2006);
T. Nishikawa and A. E. Motter, {\it ibid.} {\bf 73}, 065106(R) (2006);
J. Um, B. J. Kim, and S.-I. Lee, J. Korean Phys. Soc. {\bf 53}, 491 (2008);
S.-W. Son, B. J. Kim, H. Hong, and H. Jeong, Phys. Rev. Lett. {\bf 103}, 
228702 (2009);R. T\"{o}njes, N. Masuda, and H. Kori, Chaos {\bf 20}, 033108
(2010). 
\bibitem{synchER} J. G\'{o}mez-Garde\~{n}es, Y. Moreno, and A. Arenas,
Phys. Rev. Lett. {\bf 98}, 034101 (2007); Phys. Rev. E {\bf 75}, 066106 (2007). 
\bibitem{synchWS} H. Hong, M. Y. Choi, and B. J. Kim, Phys. Rev. E {\bf 65}, 
026139 (2002). 
\bibitem{ott} J. G. Restrepo, E. Ott, and B. R. Hunt, Phys. Rev. E {\bf 71},
036151 (2005); Chaos {\bf 16}, 015107 (2005).
\bibitem{synchSF} H. Hong, H. Park, and L.-H. Tang, Phys. Rev. E {\bf 76},
066104 (2007).
\bibitem{lattice1} H. Hong, H. Park, and M. Y. Choi, Phys. Rev. E {\bf 70},
045204(R) (2004); {\bf 72},036217 (2005). 
\bibitem{lattice2} H. Hong, H. Chate, H. Park, and L.-H. Tang, Rhys. Rev.
Lett. {\bf 99}, 184101 (2007). 
\bibitem{ER} P. Erd\"{o}s and A. R\'{e}nyi, Publicationes Mathematicae Debrencen
{\bf 6}, 290 (1959).  
\bibitem{WS} D. J. Watts and S. H. Strogatz, Nature (London) {\bf 393}, 
440 (1998).
\bibitem{inter1} S. V. Buldyrev, R. Parshani, G. Paul, H. E. Stanley, 
and S. Havlin, Nature {\bf 464}, 1025 (2010).
\bibitem{inter2} For more examples, S. Havlin, N. A. M. Araujo, 
S. V. Buldyrev, C. S. Dias, R. Parshani, G. Paul, and H. E. Stanley, 
arXiv:1012.0206v1. 
\bibitem{inter3} R. Parshani, S. V. Buldyrev, and S. Havlin, Phys. Rev. Lett.
{\bf 105}, 048701 (2010);  Proc. Natl. Aca. Sci. (USA) {\bf 108}, 1007 (2011).
\bibitem{jo} H. -H. Jo, S. K. Baek, and H. -T. Moon, Physica A {\bf 361}, 534 (2006).
\bibitem{inter4} P. A. Robinson, C. J. Rennie, and D. L. Rowe, Phys. Rev. E
{\bf 65}, 041924 (2002); J. D. Victor, J. D. Drover, M. M. Conte, and
N. D. Schiff, Proc. Natl. Aca. Sci. (USA) (in press) (doi:10.1073/pnas.1012168108).  
\bibitem{RG} O. Kogan, J. L. Rogers, M. C. Cross, and G. Refael, Phys. Rev. E
{\bf 80}, 036206 (2009); T. E. Lee, G. Rafael, M. C. Cross, O. Kogan, and
J. L. Rogers, {\it ibid.} {\bf 80}, 046210 (2009). 
\end{thebibliography}
\end{document}